\documentclass[journal]{IEEEtran}
\usepackage{stmaryrd}
\usepackage{amsmath}
\usepackage{amsthm}
\usepackage{xfrac}
\usepackage{amssymb}
\usepackage{mathabx}
\usepackage{psfrag}
\usepackage{graphicx}
\usepackage{epstopdf}
\usepackage{todonotes}
\usepackage{cite}
\usepackage{algorithmic}
\usepackage{algorithm}
\usepackage{svg}
\usepackage{mathtools}
\usepackage{hyperref}
\usepackage{hyphenat}
\usepackage{booktabs}
\usepackage{accents}
\usepackage{bbm}
\usepackage{balance}
\newlength{\dhatheight}

\usepackage{float}
\usepackage{multirow}
\usepackage{xparse,mathtools}

\begin{document}
%
% paper title
% Titles are generally capitalized except for words such as a, an, and, as,
% at, but, by, for, in, nor, of, on, or, the, to and up, which are usually
% not capitalized unless they are the first or last word of the title.
% Linebreaks \\ can be used within to get better formatting as desired.
% Do not put math or special symbols in the title.

\title{
A Novel Interference Minimizing Waveform for Wireless Channels with Fractional Delay: Inter-block Interference Analysis}
\author{Karim A. Said, A. A. (Louis) Beex, Elizabeth Bentley, and Lingjia Liu
\thanks{K. Said, A. A. Beex and L. Liu are with Wireless@Virginia Tech, the Bradley Department of ECE at Virginia Tech, Blacksburg, VA. E. Bentley is with the Information Directorate of Air Force Research Laboratory, Rome NY. }
}

% 
% *** Note that you probably will NOT want to include the author's ***
% *** name in the headers of peer review papers.                   ***
% You can use \ifCLASSOPTIONpeerreview for conditional compilation here if
% you desire.

% If you want to put a publisher's ID mark on the page you can do it like
% this:
%\IEEEpubid{0000--0000/00\$00.00~\copyright~2015 IEEE}
% Remember, if you use this you must call \IEEEpubidadjcol in the second
% column for its text to clear the IEEEpubid mark.

% use for special paper notices
%\IEEEspecialpapernotice{(Invited Paper)}

% for Transactions on Magnetics papers, we must declare the abstract and
% index terms PRIOR to the title within the \IEEEtitleabstractindextext
% IEEEtran command as these need to go into the title area created by
% \maketitle.
% As a general rule, do not put math, special symbols or citations
% in the abstract or keywords.

% make the title area
\maketitle
\begin{abstract}
In the physical layer (PHY) of modern cellular systems, information is transmitted as a sequence of resource blocks (RBs) across various domains with each resource block limited to a certain time and frequency duration. In the PHY of 4G/5G systems, data is transmitted in the unit of transport block (TB) across a fixed number of physical RBs based on resource allocation decisions. This simultaneous time and frequency localized structure of resource allocation is at odds with the perennial time-frequency compactness limits. Specifically, the band-limiting operation will disrupt the time localization and lead to inter-block interference (IBI). The IBI extent, i.e., the number of neighboring blocks that contribute to the interference, depends mainly on the spectral concentration properties of the signaling waveforms.
Deviating from the standard Gabor-frame based multi-carrier approaches which use time-frequency shifted versions of a single prototype pulse, the use of a set of multiple mutually orthogonal pulse shapes-that are not related by a time-frequency shift relationship-is proposed. We hypothesize that using discrete prolate spheroidal sequences (DPSS) as the set of waveform pulse shapes reduces IBI. Analytical expressions for upper bounds on IBI are derived as well as simulation results provided that support our hypothesis.
\end{abstract}

% Note that keywords are not normally used for peerreview papers.
%\begin{IEEEkeywords}
%Short block transmissions, Inter-block Interference, guard intervals
%\end{IEEEkeywords}

\section{Introduction}
Orthogonal frequency division multiplexing (OFDM) has been selected as the physical layer waveform for the 5G NR standard, a choice influenced mainly by considerations of maturity and backwards compatibility \cite{lien20175g}. However, there are many technical concerns regarding OFDM’s long-term sustainability mainly due to its inadequacy in high mobility scenarios \cite{Wang_2006}. In addition, OFDM's spectrum has high out-of-band (OOB) emissions which can cause significant severe interference to systems operating in adjacent frequency bands \cite{huang2015out}. 
This has motivated many efforts to investigate novel waveforms to supplant OFDM \cite{5753092}. A candidate waveform rising in popularity is Orthogonal Time Frequency Signaling (OTFS) where  information is encoded in the delay-Doppler (DD) domain \cite{hong2022delay}.  
Other DD modulation waveforms have been proposed in the literature \cite{lin2022orthogonal}. However, in some works it is argued that OTFS is a precoded version of OFDM \cite{zemen2018iterative}.

OTFS has a number of benefits, including power uniformity across symbols and channel invariance \cite{7925924}.
Nevertheless, OTFS has its own challenges such as its susceptibility to fractional Doppler \cite{9456029} and potentially fractional delay which makes cyclic prefixes corresponding to integer channel tap lengths invalid. 
One of OTFS's most celebrated advantages is the sparse structure of its equivalent channel matrix which helps in reducing the equalization complexity \cite{7993849}. However, this sparsity rests on the assumption that  delay and Doppler of the channel paths are integers when measured in units of samples and cycles/frame (normalized units), which is an unrealistic assumption. The fractional Doppler limitation is widely acknowledged in the OTFS existing literature and its impact on channel estimation accuracy and equalization complexity. From the point of view of channel estimation, works such as \cite{8671740} and \cite{9473532} study the impact of fractional Doppler on channel estimation accuracy. As a consequence, wide guard overhead regions are required to mitigate data to pilot interference and maintain channel estimation integrity. Machine learning based approaches have been used as an attempt to circumvent channel estimation altogether \cite{9745801, zhou2020learning}. From the point of view of equalization, the channel matrix sparsity advantage is lost in the presence of fractional Doppler to higher equalization computational complexity \cite{zou2021low,tusha2023low}. 

By analogy, fractional delay  presents similar problems for single carrier (SC) waveforms.  SC waveforms have been considered recently in 5G application scenarios such as massive machine type of communication (mMTC)  \cite{hu2019low} and ultra-reliable low latency communications (URRLC) \cite{tusha2019single}. SC waveforms rely on time-domain equalizers to combat inter-symbol interference (ISI), which can only handle multi-path delay taps that are integer multiples of the sample period \cite{yilmaz2022control}. The effect of fractional delay on OFDM systems has also been discussed in works such as \cite{7041655,7574382,sahin2013investigation}.

Most aforementioned works concern \textit{intra}-frame effects of fractional delay and Doppler. On the other hand, not much attention is paid to \textit{inter}-frame effects. For example, fractional delay can cause leakage between OTFS frames that extends beyond the nominal CP length. In a more general setting where information is transmitted as a sequence of blocks (OTFSs frame, OFDM symbol or single carrier block of samples) in time, fractional delay can cause Inter-block Interference (IBI) of considerable magnitude that can have an impact on symbol error rate performance as we show in our work.  

%Fractional delay and Doppler are not fundamental effects related to the channel, rather they are caused by the time-frequency concentration of the signaling waveform. For example, fractional Doppler is a result of sampling the frequency spectrum of a rectangular time domain pulse shape which is used in nominal OFDM and OTFS (MC-OTFS) \cite{7925924}. According to the Heisenberg uncertainty principle, high spectral tails can be tempered using a tapered time domain window shape. Many window shaping techniques exist in the literature to lower spectral lobes \cite{6521077} and improving performance in general \cite{rugini2006low}. However, what can be achieved is a tradeoff between spectral confinement and time confinement. In other words, using pulse shaping to lower the effect of fractional Doppler will accentuate the effect of fractional delay spread. Furthermore, in the discrete setting, pulse shaping can lead to loss of orthogonality which will affect the conditioning of the equivalent channel matrix or even lead to singularity \cite{michailow2014generalized}. % OQAM still preserves orthogonality.

Given this context, our work analyzes the effect of fractional delay on existing waveforms in terms of IBI and presents a novel waveform that can minimize IBI for a minimum sacrifice in resource utilization. Most existing waveforms such as OFDM, FBMC and OTFS can be classified under the category known as Gabor frames \cite{sahin2013survey,strohmer2001approximation}. Gabor frames consist of a set of waveforms that are time and frequency shifted versions of a \textit{single} prototype pulse shape. In this work, we propose using (a set of) \textit{multiple} mutually orthogonal pulse shapes that are not related by a time-frequency shift relationship. The pulse shape set is comprised of discrete prolate spheroidal sequences (DPSS) \cite{slepian1978prolate} for which we demonstrate its merit in terms of very low IBI. 

The main contributions of this work can be divided into the following:
\begin{itemize}
    \item A mathematical framework for quantifying the effect of fractional time or frequency shifts on the energy spread for arbitrary waveforms. In this work we focus mainly on fractional time shift but the framework is applicable to fractional shifts in frequency as well.
    \item Upper bounds on inter-block interference for arbitrary signaling waveform. 
    \item A DPSS-based signaling waveform and theoretical justification for its significantly lower IBI compared to other domains. 
\end{itemize}
 \section{System Model}

%The effect of a general linear time-variant channel on a signal consists of a super-position of multiple copies of the signal each affected by a delay and Doppler shift corresponding to a spatial propagation path \cite{bello1963characterization}. For $P$ paths, the channel effect on a signal $x(t)$ to produce $y(t)$ can be described by \eqref{analog_channel}
%\begin{equation}\label{analog_channel}
%y(t)=\sum_{p=0}^{P-1}h_px(t-\tau_p)e^{j2\pi \nu_p t}+n(t)
%\end{equation}
%where $h_p$, $\tau_p$, $\nu_p$ are the gain, delay and Doppler of the $p$-th path respectively, and $n(t)$ is the noise signal. Alternatively, the input-output relationship can be described by a convolution with a time-varying impulse response (TV-IR) in \eqref{tv_conv}
%\begin{equation}\label{tv_conv}
%y(t)=\int_{t-\tau_{max}}^{t}h(t,t-\tau)x(\tau)d\tau
%\end{equation}

We adhere to a matrix framework for representing the discrete time input-output relations of operations at the transmitter, receiver and channel effects.% For signals of finite bandwidth $B$, an equivalent discrete-time channel treatment is made possible \cite{kadous2001optimal} where time and delay are discretized as $t=nT_s, \tau=mT_s$ in \eqref{tv_conv}, where $T_s=1/B$. The discrete time version of \eqref{tv_conv} can be written in matrix form \eqref{tx_mod_ch}.
Without loss of generality, at the transmitter a frame of information symbols $\mathbf{i} \in \mathbb{C}^{I \times 1}$ modulates a set of waveforms to generate samples in the time domain represented by vector $\mathbf{x} \in \mathbb{C}^{N}$ where $\mathbf{x}=\mathbf{O}\mathbf{i}$ and $\mathbf{O} \in \mathbb{C}^{N \times I}$.
After undergoing the channel effects represented by a time-varying impulse response matrix $\mathbf{H}\in \mathbb{C}^{N\times N}$, vector $\mathbf{y}$ is acquired at the receiver:
\begin{equation}\label{tx_mod_ch}
\mathbf{y}=\mathbf{H}\mathbf{x}+\mathbf{n}
\end{equation}
where $\mathbf{n}$ is the noise vector.
A matched filtering operation is applied by correlating with the transmit waveform set (or its co-set) for bi-orthogonal schemes:
\begin{equation}\label{eq_io}
\begin{split}
\mathbf{z}&=\mathbf{O}^H\mathbf{H}\mathbf{O}\mathbf{i}+\mathbf{O}^{H}\mathbf{n}\\
&=\mathbf{H}_{eq}\mathbf{i}+\mathbf{O}^H\mathbf{n}
\end{split}
\end{equation}

Discretizing in time, for a finite stream of $LK$ symbols, $x(t)$ in \eqref{pulse_train} can be written in terms of a matrix-vector product: 
\begin{equation}\label{mod_io_0}
\mathbf{x}=\left(\mathbf{I}_L\otimes  \mathbf{Q} \right)\mathbf{i}
\end{equation}
where $\mathbf{i}=[i_0,..,i_{LK-1}]^T$, $\mathbf{Q}=[\mathbf{q}_0,\mathbf{q}_2,..\mathbf{q}_{K-1}], \mathbf{q}_i \in \mathbb{C}^{M'\times 1}$ where $M'\geq K$, and $\mathbf{I}_L$ is an identity matrix of size $L\times L$.

In OFDM and similar modulation schemes such as generalized OFDM (G-OFDM) \cite{michailow2014generalized} and filter-bank multi-carrier (FBMC) \cite{5753092} the domain with coordinates $0,..,K-1$ is frequency. Vector $\mathbf{q}_i$ represents a modulated version of a single prototype pulse shape $\mathbf{q}$; a category of modulation schemes known as Gabor frames \cite{hlawatsch2011wireless}, where $\mathbf{q}_k=\textbf{diag}(\mathbf{f}_k)\mathbf{q} $ and $[\mathbf{f}_k]_n=e^{j\frac{2\pi km}{M}}, m=0,..,M'-1$.

Fig. \ref{TD_TV_channel} shows the interaction between a channel of maximum delay spread $\tau_{max}$, and a pulse shape consisting of two sub-blocks,   $\mathbf{p}= [\mathbf{g}^T,\mathbf{q}^T]^T \in \mathbb{C}^{M \times 1},  \mathbf{g} \in \mathbb{C}^{\tau_{max} \times 1}, M = M'+\tau_{max}$. %The block $\mathbf{g}$ is to perform a function similar to the spacing between pulses in  \eqref{pulse_train}. 
Now \eqref{mod_io_0} changes to:
\begin{equation}\label{mod_io}
\mathbf{x}=\left(\mathbf{I}_L\otimes  \mathbf{P} \right)\mathbf{i}
\end{equation}
where $\mathbf{P}=[\mathbf{p}_0,\mathbf{p}_2,..\mathbf{p}_{K-1}].$

%The channel is depicted in Fig. \ref{TD_TV_channel} by the thin and tall parallelogram having a small side of length $\tau_{max}$. An input consisting of a sequence of two packets $\mathbf{p}$ is shown on the top and the corresponding output is shown on the left.
%\begin{figure}
%\centering 
%\includegraphics[width=0.8\linewidth]{Integer_tap_channel_subblock_interaction3.png}
%\caption{Matrix representing Time-varying impulse response (TV-IR) of channel with integer delay taps depicted by diagonal parallelogram maps input blocks to output blocks. (Top) a sequence of two input blocks, each block consisting of sub-blocks $\mathbf{g}$ and $\mathbf{p}$. (Left) Corresponding received block for each of the transmitted sub-blocks.} \label{TD_TV_channel}
%\end{figure}
%Dashed rectangle braces define the borders of the elements of a matrix. Matrices with yellow braces define the contribution of each block in the input to the corresponding block in the output. The matrix with red braces accounts for contributions from the first block in the input to the second block in the output, i.e., inter-block interference (IBI).

A common strategy to eliminate this form of IBI is to set sub-block $\mathbf{g}$ to zero, where $\mathbf{g}$ is called a zero prefix (ZP) and ignore the corresponding sub-block in the output. Another strategy  is to set $\mathbf{g}= [p_{N-\tau_{max}+1}:p_N]^T$ where $\mathbf{g}$ is called a cyclic prefix (CP) where the submatrix represented by the blue triangle effectively translates to the upper right corner of  yellow border matrices as depicted by the faded blue triangles in Fig. \ref{TD_TV_channel}.

Substituting \eqref{mod_io} into \eqref{eq_io} and referring back to the aforementioned objective of shaping $\mathbf{H}_{eq}$ to be close to a diagonal structure, we can see that using a ZP makes it possible to obtain a \textit{block-diagonal} structure.
\begin{equation}\label{eq_io_block}
\begin{split}
\mathbf{z}&=\left(\mathbf{I}_L\otimes  \mathbf{P} \right)^H\mathbf{H}\left(\mathbf{I}_L\otimes  \mathbf{P} \right)\mathbf{i}+\left(\mathbf{I}_L\otimes  \mathbf{P} \right)^H\mathbf{n}\\
&=\mathbf{H}_{eq}\mathbf{i}+\left(\mathbf{I}_L\otimes  \mathbf{P} \right)^H\mathbf{n}
\end{split}
\end{equation}
where $\mathbf{H}_{eq}$ is block diagonal matrix as in \eqref{blk_mat}, and $\mathbf{n}\in \mathbb{C}^{N \times 1}$ is a AWGN noise vector.
\begin{equation} \label{blk_mat}
\mathbf{H}_{eq}= \text{blkdiag}(\mathbf{H}_{0,0},\mathbf{H}_{1,1},..,\mathbf{H}_{L-1,L-1})
\end{equation}
As a result, \eqref{eq_io_block} can be separated into smaller sets of equations:
 \begin{equation}\label{eq_io_one_block}
\begin{split}
\mathbf{z}_l=\mathbf{P}^H\mathbf{H}_{l,l}\mathbf{P}\mathbf{i}_l+\mathbf{P}^H\mathbf{n}_l, \quad l=0,..,L-1
\end{split}
\end{equation}
 where $\left[\mathbf{H}_{l,l'}\right]_{m,m'}=\left[\mathbf{H}\right]_{lM+m,l'M+m'}, m,m'=0,..,M-1$, $\left[\mathbf{n}_{l}\right]_{m}=\left[\mathbf{n}\right]_{lM+m}, m=0,..,M-1$, $\left[\mathbf{i}_{l}\right]_{m}=\left[\mathbf{i}\right]_{lM+m}, m=0,..,M-1$ and $\tau_{max}$ is the maximum delay.
 
 The significant advantage of such a block channel structure is that, through proper choice of $\mathbf{g}$, equalization can be done on a block-by-block level and that greatly reduces complexity. This inspires our strategy to design a waveform where the block length can be made as small as possible. In doing so, we must address the consequences of using small block lengths on the manifestation of channel effects related to delay spread.

\subsection{Discrete Doubly Dispersive Channel Model}

In a typical communication system, time and bandwidth constraints are simultaneously enforced; transmit filters strictly limit the signal bandwidth, at the receiver side, the received signal is forced to be limited when evaluating its inner product against a finite extent reference block.
The limit on signaling bandwidth and time extent of a signaling block induces a discrete time channel matrix representation of the time-varying impulse response (TV-IR). For a channel with $P$ discrete specular paths:
\begin{equation}\label{ch_matrix}
\begin{split}
\mathbf{H}&=\sum_{p=0}^{P-1}\mathbf{H}_{\nu\tau_p}=\sum_{p=0}^{P-1}h_p\mathbf{D}_{\nu_p}\mathbf{H}_{\tau_p}
\end{split}
\end{equation}
where $\left[\mathbf{H}_{\nu_p}\right]_{l,k}=e^{j2\pi l\nu_p}\delta[l-k]$ represents the Doppler modulation effect for normalized Doppler frequency $\nu_p$, $[\mathbf{H}_{\tau_p}]_{l,k}=\frac{\sin\pi(l-k-\tau_p)}{\pi(l-k-\tau_p)}$ is the delay effect for normalized delay $\tau_p$ for the $p$-th path respectively and $h_p$ is the path gain.

Now we analyze the structure of the matrix $\mathbf{H}$ by looking at the structure of the individual summand matrices $\mathbf{H}_{\nu\tau_p}$. Each summand matrix is the product of a (main) diagonal matrix $\mathbf{D}_{\nu_p}$ and Toeplitz matrix $\mathbf{H}_{\tau_p}$. Matrix $\mathbf{H}_{\tau_p}$ will have exactly one (sub) diagonal if $\tau_p$ is an integer, otherwise it will be a full matrix. As a result, the product matrix inherits the diagonal extent of $\mathbf{H}_{\tau_p}$ (spanning the full matrix while decaying in the anti-diagonal direction for non-integer delays) but loses the property of being Toeplitz. An illustration is shown in Fig. \ref{matrix_product}. 
Thus, the thin parallelogram depiction in Fig. \ref{TD_TV_channel} is true only if all (normalized) path delays are integers. As a consequence, CP or ZP approaches will not completely eliminate inter-block interference and \eqref{eq_io_one_block} is modified as follows to include an IBI term $\boldsymbol\beta_l$:
\begin{equation}\label{eq_io_one_block_IBI}
\begin{split}
\mathbf{z}_l&=\mathbf{P}^H\mathbf{H}_{l,l}\mathbf{P}\mathbf{i}_l+\boldsymbol\beta_l+\mathbf{P}^H\mathbf{n}_l\\
\end{split}
\end{equation}
where 
\begin{equation}\label{beta_i}
\begin{split}
\boldsymbol \beta_l &= \sum_{j=0,j\neq l}^{L-1}\mathbf{P}^H\mathbf{H}_{l,j}\mathbf{P}\mathbf{i}_j= \sum_{p=0}^{P-1}h_p\sum_{j=0,j\neq l}^{L-1}\boldsymbol \Lambda_{lj}(p)
\end{split}
\end{equation}
where $\boldsymbol \Lambda_{lj}(p)=\mathbf{P}^H\mathbf{D}_{l,l}(\nu_p)\mathbf{H}_{l,j}(\tau_p)\mathbf{P}$.
%\begin{figure}
%\centering 
%\includegraphics[width=1\linewidth]{Matrix_bandedness_cropped.png}
%\caption{Delay and Doppler factor matrices for a doubly dispersive channel: (Left) Doppler matrix, (Center) delay matrix, (Right) product of the two matrices.}\label{matrix_product}
%\end{figure}
 \subsection{Impact of waveform choice on IBI in a purely delay-dispersive channel}
Let $\tilde{\mathbf{P}} \in \mathbb{C}^{N \times N}$ be a unitary matrix such that $\mathbf{P} \in \mathbb{C}^{N \times L}$  comprises the first $L$ columns of $\tilde{\mathbf{P}}$ where $L\leq N$. Substituting $\tilde{\mathbf{P}}\tilde{\mathbf{P}}^H$ into $\boldsymbol \Lambda_{lj}(p)$ enables the separation of the Doppler effect and delay effect into one distinct matrix for each.
\begin{equation}\label{separation}
\begin{split}
\boldsymbol \Lambda_{lj}(p)&=\mathbf{P}^H\mathbf{D}_{l,l}(\nu_p)\tilde{\mathbf{P}}\tilde{\mathbf{P}}^H\mathbf{H}_{l,j}(\tau_p)\mathbf{P}=\tilde{\boldsymbol \Lambda}^{\nu}_{l}(p)\tilde{\boldsymbol \Lambda}^{\tau}_{lj}(p)
\end{split}
\end{equation}
Since the \textit{span} of IBI is only dependent on the delay dispersion, i.e., $\tilde{\boldsymbol \Lambda^{\tau}}_{ij}(p)$, we focus our analysis on purely delay-dispersive channels. In a purely delay-dispersive channel, the contribution of the $j$-th input to IBI affecting block $l$ due to interacting with the $p$-th channel becomes:
\begin{equation}\label{single_path_IBI}
\begin{split}
\boldsymbol \beta_{l}(p)=h_p\sum_{j=0,j\neq l}^{L-1}\boldsymbol \Lambda_{lj}^{\tau}(p)\mathbf{i}_j
\end{split}
\end{equation}
where $\boldsymbol \Lambda^{\tau}_{lj}(p)=\mathbf{P}^H\mathbf{H}_{l,j}(\tau_p)\mathbf{P} \in \mathbb{C}^{K\times K}$, and
\begin{equation}\label{IBI_per_wvfrm}
[\boldsymbol \beta_{l}(p)]_r=h_p\sum_{j=0,j\neq l} \sum_{k=0}^{K-1}\boldsymbol [\Lambda_{lj}^{\tau}(p)]_{r,k}[\mathbf{i}_j]_k
\end{equation}

The $r,s$ element of the delay factor matrix, i.e., $[\boldsymbol \Lambda_{ij}^{\tau}(p)]_{r,s}$ is given by \eqref{delay_factor_matrix_element}
\begin{equation}\label{delay_factor_matrix_element}
\begin{split}
[\boldsymbol \Lambda_{lj}^{\tau}(p)]_{r,s}&=\sum_{n,m=0}^{N-1} \mathbf{p}_r^*[n] \mathbf{p}_s[m]\text{sinc}(n-m-(j-l)N-\tau_p)\\
&=\sum_{q=-N+1}^{N-1} \mathbf{c}_{rs}[q]\text{sinc}\left(q-(\tau_p+(j-l)N)\right)\\
\end{split}
\end{equation}
where $\mathbf{c}_{rs}[q]=\sum_{n=\max(-N/2+q,-N/2)}^{\min(N/2+q,N/2)}\mathbf{p}_r^*[n]\mathbf{p}_s[n-q]$, $\text{sinc}(x)=\frac{\sin \pi x}{\pi x}$. 

The second line in \eqref{delay_factor_matrix_element} is the $r-s$-th cross-correlation sequence, $\mathbf{c}_{rs}$, shifted by $\tau_p+(j-l)N$. $\mathbf{c}_{rs}$ is an index limited sequence which when shifted is convolved with a sequence that is infinite in extent for fractional values of $\tau_p$.  This elongation effect is the underlying cause of IBI that extends past the cyclic/zero prefix. To the best of our knowledge, no works exist which simplify the last line in \eqref{delay_factor_matrix_element} to an analytical closed form expression for fractional shifts $\tau_p$. Finding such an analytical expression would enable us to quantify the IBI energy which can provide us some measure of the expected degradation in SER performance. 

In what follows we pursue analytical expressions for upper bounding IBI for any waveform set choice. Towards this end, we rely on the mathematical framework that was developed in \cite{10510883}.

\subsection{Upper bounding Energy of Cross-correlation Tail due to Fractional Shift}
%\vspace{-5pt}
Rewriting \eqref{delay_factor_matrix_element} in terms of $\lbrace \mathcal{B}_W^{\tau}\rbrace r[n]$, defined by (1) in \cite{10510883} as the operation on a sequence $r[n]$ that outputs a sequence limited in frequency to half-bandwidth $W$ (normalized), scaled by $1/W$, and shifted by $0<\tau\leq 0.5$, results in
%\vspace{-4pt}
\begin{equation}\label{delay_factor_matrix_XCORR}
\begin{split}
[\boldsymbol \Lambda_{lj}^{\tau}(p)]_{r,s}&=\left\lbrace\mathcal{B}_W^{\tau_p+(j-l)N} \mathbf{c}_{rs}\right\rbrace =\left\lbrace\mathcal{B}_W^{\tau_p} \mathbf{c}_{rs}\right\rbrace[(j-l)N] \\
\end{split}
\end{equation}
In order to quantify IBI energy, we are interested in the quantity given by the LHS of \eqref{one_path_XCORR_tail_energy}, i.e., energy of the sub-sampled tail (by a factor of $N$) of the $r-s$th cross-correlation sequence. The first line of the RHS of \eqref{one_path_XCORR_tail_energy} consists of an upper bound in terms of the cross-correlation tail energy, where $\bar{E}_{-N,N}$
denotes the tail energy of the correlation sequence $\mathbf{c}_{rs}$,  according to Definition-1 in \cite{10510883} 
\begin{equation}\label{one_path_XCORR_tail_energy}
\begin{split}
\sum_{i=-\infty\neq j}^{\infty}\left\lbrace\mathcal{B}_W^{\tau_p} \mathbf{c}_{rs}\right\rbrace^2[(i-j)N] &\leq \bar{E}_{-N,N}\left(\left\lbrace\mathcal{B}_W^{\tau_p} \mathbf{c}_{rs}\right\rbrace\right) \\
&\leq \bar{E}_{-N,N}\left(\left\lbrace\mathcal{B}_W^{0.5} \mathbf{c}_{rs}\right\rbrace\right) \\
&\leq \sum_{l=0}^{4N} \left|\frac{c_{r,s}(l)}{W}\right|^2\lambda_l(1-\lambda_l)
\end{split}
\end{equation}
where $c_{r,s}(l)=\sum_{n=-2N}^{2N}\mathbf{c}_{rs}[n]s_l^{(0.5W,4N+1)}[2n]$. The inequality in the second line is based on our conjecture that: out of all fractional sample shifts, a half sample shift results in the largest tail energy. In the third line, the result of Theorem-1 in \cite{10510883} (equation (15)) is applied.
We note that the LHS of \eqref{one_path_XCORR_tail_energy} involves a computation potentially involving an infinite number of terms; when a general infinite stream of blocks across time is considered. The given bound requires only a finite number of computations that does not depend on the number of blocks transmitted across time.
\section{Quantifying IBI power in Delay Dispersive Channels}
In a delay-dispersive channel consisting of $P$ paths, we can find the IBI energy affecting the $r$-th waveform by averaging \eqref{IBI_per_wvfrm} over information symbols which are assumed to be unit variance i.i.d. Without loss of generality, we start by setting $l=0$ in \eqref{beta_i} and evaluating the following:

\begin{equation}\label{IBI_multipath}
\begin{split}
E_{r}^{IBI}&= \mathbb{E}\left\lbrace\left|\sum_{p=0}^{P-1}h_p\beta_0(p)\right|^2\right\rbrace\\
&= \mathbb{E}\left\lbrace\left|\sum_{p=0}^{P-1}h_p\sum_{{\substack{j=-\infty, \\ j\neq 0}}}^{\infty}\sum_{s=0}^{K-1}\boldsymbol [\Lambda_{0j}^{\tau}(p)]_{r,s}[\mathbf{i}_j]_s\right|^2\right\rbrace\\
&= \sum_{p=0}^{P-1}|h_p|^2\sum_{{\substack{j=-\infty, \\ j\neq 0}}}^{\infty}\sum_{s=0}^{K-1} \left|\left\lbrace\mathcal{B}_W^{\tau_p} \mathbf{c}_{rs}\right\rbrace[jN']\right|^2\\
&= \sum_{p=0}^{P-1}|h_p|^2\sum_{{\substack{j=-\infty, \\ j\neq 0}}}^{\infty}\sum_{s=0}^{K-1} \left|\left\lbrace\mathcal{B}_W^{\Delta \tau_p} \mathbf{c}_{rs}\right\rbrace[jN'+\left\lfloor\tau_p\rfloor\right]\right|^2\\
\end{split}
\end{equation}
The simplification from the second to third line is due to $\mathbb{E}\lbrace [\mathbf{i}_j]_s[\mathbf{i}_{j'}]_{s'}\rbrace=\delta(j-j',s-s')$ , $\mathbb{E}\lbrace h_ph_{p'}\rbrace=|h_p|^2\delta(p-p')$, and $\Delta\tau_p\triangleq \left(\tau_p-\lfloor\tau_p\rfloor\right)$.

Appending waveforms with a guard prefix of length $g\geq\lfloor \tau_p \rfloor,  \forall p$, changes \eqref{IBI_multipath} as follows:
\begin{equation}\label{IBI_split}
\begin{split}
&E_{r}^{IBI}\\
&=\sum_{p=0}^{P-1}|h_p|^2\sum_{s=0}^{K-1}\sum_{\substack{j=-\infty, \\\neq 0}}^{\infty} \left|\left\lbrace\mathcal{B}_W^{\Delta \tau_p} \mathbf{c}_{rs}\right\rbrace[j(N'+g)+\left\lfloor\tau_p\rfloor\right]\right|^2\\
&\leq \sum_{p=0}^{P-1}|h_p|^2\sum_{s=0}^{K-1}\sum_{{\substack{j=-\infty, \\ \neq[-(N'+g),(N'+g)]}}}^{\infty} \left|\left\lbrace\mathcal{B}_W^{\Delta \tau_p} \mathbf{c}_{rs}\right\rbrace[j+\left\lfloor\tau_p\rfloor\right]\right|^2\\
&= \sum_{\substack{p=0,\\ \Delta \tau_p>0}}^{P-1}|h_p|^2\sum_{s=0}^{K-1}E_{-(N'+g-\lfloor\tau_{p}\rfloor),(N'+g-\lfloor\tau_{p}\rfloor)}\left(\left\lbrace\mathcal{B}_W^{\Delta\tau_p} \mathbf{c}_{rs}\right\rbrace\right)\\
\end{split}
\end{equation}
We note that the guard prefix results in $\bar{E}_{-N',N'}\left(\left\lbrace\mathcal{B}_W^{0.5} \mathbf{c}_{rs}\right\rbrace\right)=0$ when $\Delta\tau_p=0$, hence the restriction in the last line of the sum indices for fractional delay paths.

To bound the total IBI for a subset of the waveforms across the range $0,..,\eta K-1$,
\begin{equation} \label{E_IBI_ubound}
\begin{split}
E^{IBI}
&\leq E^{IBI}_r\\
&=\sum_{\substack{p=0,\\ \Delta \tau_p>0}}^{P-1}|h_p|^2\sum_{r=0}^{\eta K-1}\sum_{s=0}^{\eta K-1} \sum_{l=0}^{4N_p} \left|\frac{c_{r,s}(l;N_p)}{W}\right|^2\lambda_l(1-\lambda_l)\\
\end{split}
\end{equation}
where $N_p = N'+g-\lfloor\tau_p\rfloor$,  $c_{r,s}(l;N_p)=\sum_{n=-2N_p}^{2N_p}\mathbf{c}_{rs}[n]s_l^{(0.5W,4N_p+1)}[2n]$.
Finally, for unit symbol energy, we can obtain a signal-to-inter-block interference (S2IBI) lower bound by taking the reciprocal of \eqref{E_IBI_ubound}.

%Using the result of Theorem-2, we derive an upper bound expression that is computable since it only involves finite sums. 
%\begin{equation}\label{E_IBI_ubnd}
%\begin{split}
%E_{l'}^{IBI}& \leq \sum_{p=0,\Delta \tau_p>0}^{P-1}|h_p|^2\sum_{s}^{}\bar{E}_{-N',N'}\left(\left\lbrace\mathcal{B}_W^{0.5} \mathbf{C}_{rs}\right\rbrace\right)\\
%&\leq \text{max}(|h_p|^2)\sum_{l=-L/2}^{L/2}\bar{E}_{-N',N'}(s_{l'},s_l)\\
%\end{split}
%\end{equation}
%The last line in \eqref{E_IBI_ubnd} is the sum of cross-correlation tail energy upper bounds of the $l'$-th sequence and every other sequence in the set.

\section{Results}

We numerically evaluate performance in terms of BER vs. SNR across three waveforms comprised of orthonormal bases (ONB) in the following domains: time domain (TD), frequency domain (FD), and Prolate spheroidal domain (PS). The BER curves are generated across different values for the resource utilization percentage  $\eta\%$. In addition, we evaluate signal-to-IBI (S2IBI) to ascertain its role as the underlying differentiating factor in BER performance between the waveforms in the different domains. Our hypothesis is that IBI due to fractional delay taps is a significant factor which is grossly ignored in models assuming integer delay taps.
Our results are based on 100 frame realizations, each frame consisting of 21 blocks, each of length $N=129$ time domain samples.  The $N$ sample block is comprised of $M=\eta N$ sub-waveforms modulated by QPSK symbols, where $\eta$ varies across the values in the range $[0.9 ,1]$ reflecting the percentage of nulled signaling dimensions as explained in Section II. 
A delay-dispersive channel with delay spread spanning $[0,\tau_{max}=16]$ samples, i.e., $1/8$th the block size and following an exponential delay profile is considered to act on the frame. We consider two channels of varying severity, as controlled by the rate of decay of the exponential delay profile: our mild channel has tap gains decaying according to $e^{-0.5 n}$ (with uniformly random phase), and our severe channel has tap gains decaying according to $e^{-0.05n}$ where $n=0,..,15$ for the integer tap case, and $n=0,0.1,0.2,..,15$ for the fractional tap case.
The frame structure is given by \eqref{sim_frame}
\begin{equation}\label{sim_frame}
\mathbf{f}=[\mathbf{0}_D^T,\mathbf{p}_{-10}^T,..,\mathbf{0}_D^T,\mathbf{p}_{0}^T,\mathbf{0}_D^T,..,\mathbf{0}_D,\mathbf{p}_{10}^T]
\end{equation}
where $\mathbf{p} = \mathbf{O}\mathbf{d}$, $\mathbf{O}\in \mathbb{C}^{N\times M}$ is the signaling basis, $\mathbf{d}\in \mathbb{C}^{M \times 1}$ is a vector of QPSK symbols, and $\mathbf{0}_D \in \mathbb{R}^{D\times 1}$ is an all zero vector where $D=\tau_{max}=16$ .

Figure \ref{S2IBI_exp_pnt5_plots} shows the variation of S2IBI (dB) with $\eta$ for the mild channel case. For the integer tap case shown in sub-figure \ref{S2IBI_exp_pnt5_plots}-(b), S2IBI is very high, which is a result of the fact that the ZP length is equal to $\tau_{max}$ and thus the delay spread is fully encompassed leading to zero IBI. 

In sub-figure \ref{S2IBI_exp_pnt5_plots}-(a), the fractional tap channel case is shown. At $\eta=1$ all three signaling domains have the same S2IBI $~28.7$ dB which can be explained by the fact that the three waveforms are \textit{complete} ONBs and thus are equivalent when $\eta=1$. 

For $\eta<1$, S2IBI rises rapidly for the PS domain waveform at a rate of $~12$ dB per $0.02$ reduction in $\eta$ with an S2IBI reaching up to $110$ dB. On the other hand, TD and FD S2IBI rise at a much slower rate, with an S2IBI reaching up to $30$ dB for TD, and $31$ dB for FD at $\eta=0.9$. Lower bounds for S2IBI based on the IBI upper bound given in \eqref{E_IBI_ubound} are shown by black markers on top of dashed lines with the same color as the bounded S2IBI for a given domain. In general the bound is not very tight, however, it closely follows the general trend of the true S2IBI shown in solid lines. We note that IBI is effectively the energy of a \textit{sampled version} of the cross-correlation sequence tail as indicated by \eqref{one_path_XCORR_tail_energy}. This explains why the IBI upper bound is not expected to be very tight and as a consequence the S2IBI lower bound will also not be very tight.

%\begin{figure}
%\centering
%\includegraphics[width=\linewidth]{S2IBI_results_9_28_2023_exp_pnt5.png}
%\caption{Signal-to-inter-block interference (S2IBI) versus resource utilization percentage for \textbf{mild channel}.  Frequency domain signaling (FD), time domain (TD) and  prolate domain (PS) depicted by green, blue and red curves respectively. (a) Channel with fractional taps, (b) Channel with integer taps}.\label{S2IBI_exp_pnt5_plots}
%\end{figure}
\begin{figure}
\centering
\includegraphics[width=\linewidth]{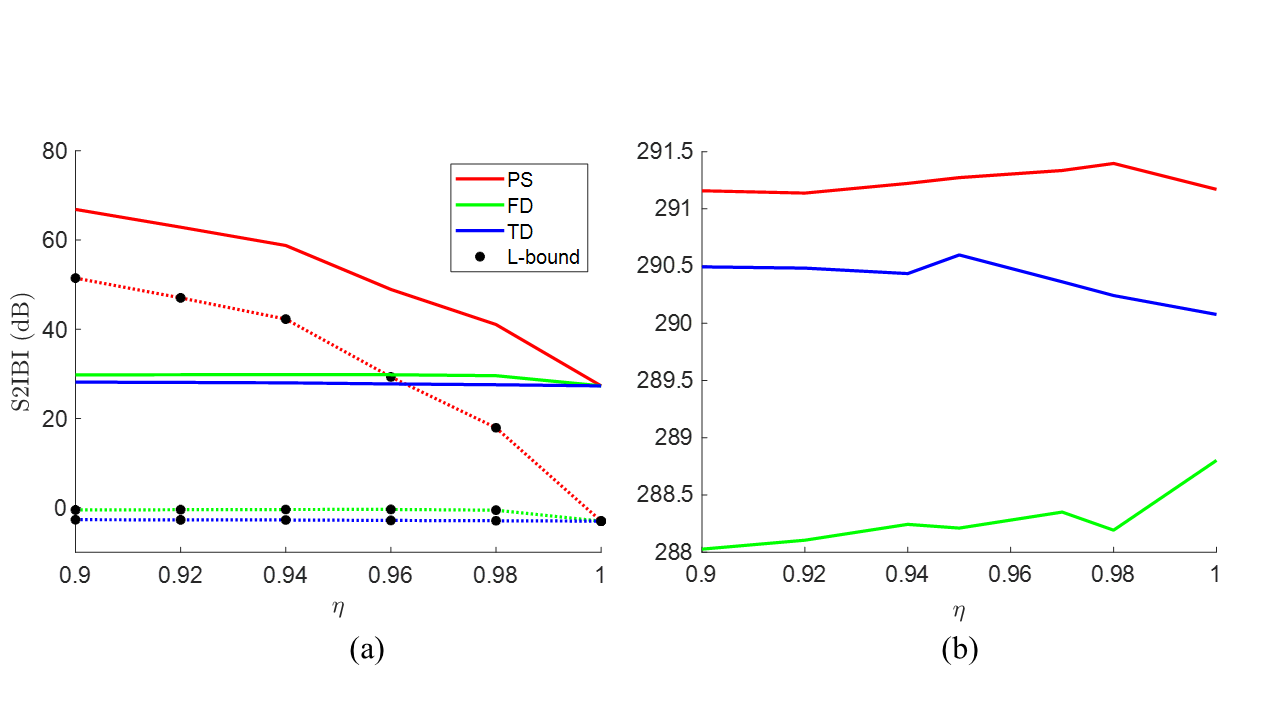}
\caption{Signal-to-inter-block interference (S2IBI) versus resource utilization percentage for \textbf{severe channel}.  Frequency domain signaling (FD), time domain (TD) and  prolate domain (PD) depicted by green, blue and red curves respectively. (a) Channel with fractional taps, (b) Channel with integer taps.}\label{S2IBI_exp_pnt05_plots}
\end{figure}
\begin{figure*}
\centering
\includegraphics[width=0.75\linewidth]{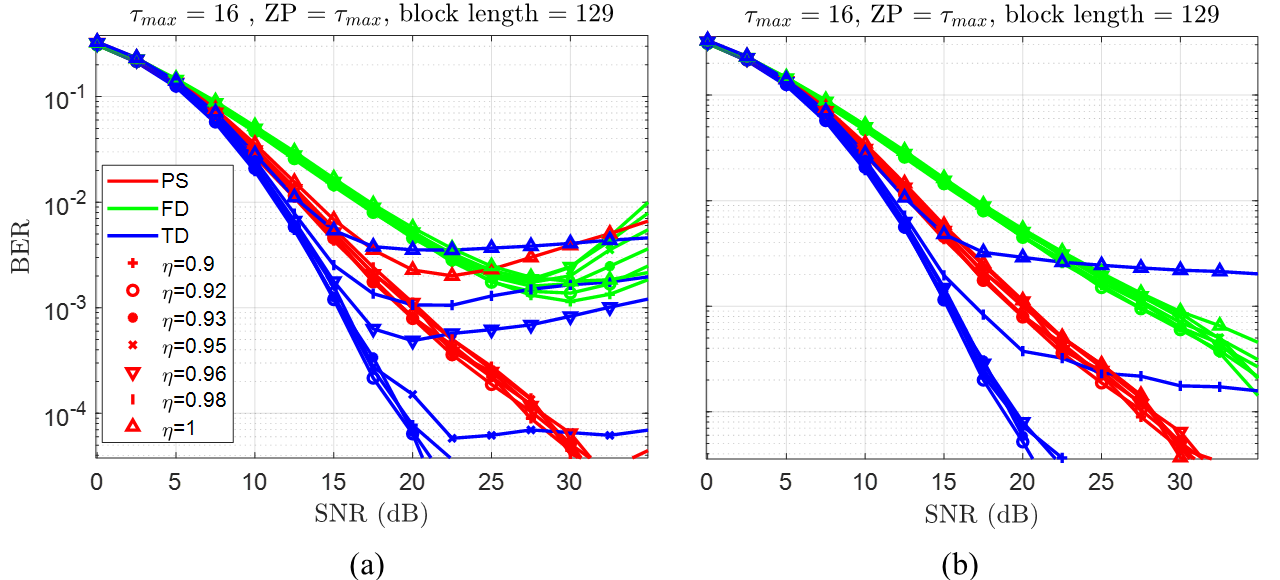}
\caption{BER vs. SNR in \textbf{mild channel} for different values of $\eta$ depicted by different markers.  Frequency domain signaling (FD), time domain (TD) and  prolate domain (PS) depicted by green, blue and red curves respectively. (a) Channel with fractional taps, (b) Channel with integer taps.}\label{BER_exp_pnt5}
\end{figure*}

Figure \ref{S2IBI_exp_pnt05_plots} shows the variation of S2IBI (dB) with $\eta$ for the severe channel case. For the integer tap case shown in sub-figure \ref{S2IBI_exp_pnt05_plots}-(b) the result is nearly the same as for the mild channel since IBI is identically 0. 
For the fractional case shown in sub-figure \ref{S2IBI_exp_pnt05_plots}-(a), at $\eta=1$ all three signaling domains have the same S2IBI $\approx26.7$ dB. For $\eta<1$, S2IBI increases rapidly for the PS domain waveform but at a lower rate compared to the mild channel case shown in Fig \ref{S2IBI_exp_pnt5_plots}-(a), reflecting the severity of the channel induced IBI.  For TD and FD S2IBI rises to $\approx27.4$ dB  and $\approx 30.6$ dB at $\eta=0.9$ for TD and FD respectively. The theoretical lower bounds are looser than the ones in Fig. \ref{S2IBI_exp_pnt5_plots} but fairly in line with the trend of the true S2IBI. 
\begin{figure*}
\centering
\includegraphics[width=0.75\linewidth]{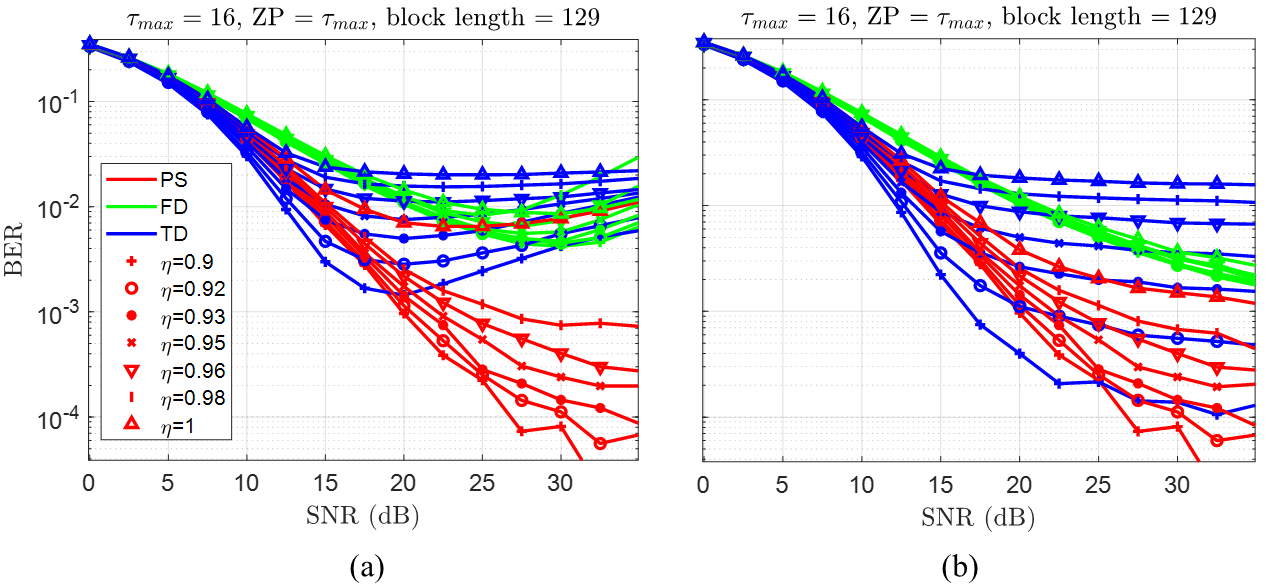}
\caption{BER v.s. SNR in \textbf{severe channel} for different values of $\eta$ depicted by different markers.  Frequency domain signaling (FD), time domain (TD) and  prolate domain (PS) depicted by green, blue and red curves respectively. (a) Channel with fractional taps, (b) Channel with integer taps.}\label{BER_exp_pnt05}
\end{figure*}
Figure \ref{BER_exp_pnt5} shows the BER performance vs. SNR across the set of resource utilization percentages $\eta=[0.9,0.92,0.93,0.95,0.96,0.98,1]$ for the mild channel case. For the fractional tap case shown in sub-figure \ref{BER_exp_pnt5}-(a), TD has error floors $\approx 3\times 10^{-3} $, $10^{-3}$, $5\times 10^{-4}$ for $\eta=1,0.98,0.96$ respectively. For values of $\eta<0.95$ TD has no visible error floor,  however the performance is different (slightly worse) than in the integer tap case. For FD, the error floor persists for all $\eta$ values ranging within $[10^{-3},2\times 10^{-3}]$. On the other hand, PS has an error floor $2\times 10^{-3}$ for $\eta=1$ since the IBI is high; the same level as FD and TD. For $\eta<1$, there are no visible error floors. For the integer tap case shown in sub-figure\ref{BER_exp_pnt5}-(b), the performance curves do not change significantly for different values of $\eta$ except for TD where error floors occur at BER $2\times 10^{-3}$ starting at $15$ dB SNR for $\eta=1$, and BER $2\times 10^{-4}$ starting at $30$ dB SNR for $\eta=0.98$. For $\eta\leq 0.98$ TD BER drops monotonically reaching down to $5\times 10^{-5}$ at $\approx20$ dB. For FD and PS, BER drops monotonically reaching down to $5\times 10^{-5}$ at $\approx30$ dB for PS, and  between $2\times 10^{-4}$ and $5\times 10^{-4}$ at $~35$ dB for FD. 

For the fractional tap channel in Fig. \ref{BER_exp_pnt05}-(a), TD has error floors ranging between slightly less than $10^{-3}$ for $\eta=0.9$ up to slightly less than $10^{-2}$ for $\eta=0.98$. We note that for TD, BER curves behave in a convex manner with a minimum at $\approx20$ dB. This can be explained by the fact that LMMSE equalization is based on a regularization factor that accounts for noise but not for IBI. From Fig. \ref{S2IBI_exp_pnt05_plots}, the S2IBI level being at $\approx 27.4$ dB makes it somewhat on the order of the noise level. As a result, for SNRs higher than $20$ dB, LMMSE is in "zero-forcing mode" leading to IBI amplification. For FD, the error floor does not go below $4\times 10^{-3}$. The convex behavior can also be seen in this case, with the minimum happening at a different point compared to TD, in agreement with the higher IBI for FD as indicated in Fig. \ref{S2IBI_exp_pnt05_plots}. For PS, an error floor is present at $\eta=1$ but not for values $\eta<1$. We also note that for PS, the performance curves for $\eta<1$ for both the integer and fractional tap model are almost identical.

In Fig. \ref{BER_exp_pnt05}-(b), just as in Fig. \ref{BER_exp_pnt5}-(b) for all three signaling domains the BER is monotonically decreasing with SNR, however the improvement rates are much slower compared to the mild channel. We note that this is unlike the case where the BER curves have a non-decreasing behavior, i.e., increasing SNR will not help, in which case we refer to this behavior as  an error floor.  FD shows almost no dependence on the choice of $\eta$, similar to the mild channel case. However, for TD and PS, lowering $\eta$ produces significant improvements in the rate of reduction of BER vs. SNR. Note that this improvement cannot be attributed to IBI since IBI is already 0 by virtue of the fact that the channel consists of integer taps. 

We note that our proposed DPSS based waveform shows its clear advantage in severe channel scenarios where IBI is prominent. However, for shorter block lengths compared to what is used in our simulation, such as in the case of ultra-reliable low-latency communications (URLLC) \cite{8047997}, the effect of IBI is expected to be more dominant even in mild channels. In addition, modulation (QPSK used in our simulation) order is expected to be a factor in amplifying the effect of IBI.

\section{Conclusion}
Inter-block interference is a problem that has its origins going back to the time-frequency concentration dichotomy. Limiting IBI can only be done at a cost in either time or bandwidth resources or in some other dimension.  
In this work, we provide strong evidence that waveforms using discrete prolate spheroidal sequences are optimal in minimizing IBI. The issue addressed has relevance beyond IBI spread as it also concerns other forms of intra-block interference. Many existing waveform designs can be thought of as consisting of micro-blocks, and thus the present analysis can be extended to address inter-waveform interference. 
Such a treatment can be key to addressing a number of pressing practical problems affecting prominent waveforms, namely fractional Doppler and fractional delay simultaneously. We plan to address such problems in our future works.

\balance

\IEEEpeerreviewmaketitle

% if have a single appendix:
%\appendix[Proof of the Zonklar Equations]
% or
%\appendix  % for no appendix heading
% do not use \section anymore after \appendix, only \section*
% is possibly needed

% use appendices with more than one appendix
% then use \section to start each appendix
% you must declare a \section before using any
% \subsection or using \label (\appendices by itself
% starts a section numbered zero.)
%

\bibliographystyle{IEEEtran}
\bibliography{IEEEabrv,references.bib}

\end{document}